\documentclass[10pt]{article}

\usepackage[numbers]{natbib}
\usepackage{microtype}
\usepackage{graphicx}
\usepackage{fullpage}
\usepackage{hyperref}
\usepackage{color}
\usepackage{url}
\usepackage{amsmath}
\usepackage{wasysym}

\hypersetup{colorlinks,breaklinks,
            citecolor=blue,
            urlcolor=blue,
            linkcolor=blue}

\begin{document}
\thispagestyle{empty}

\hrule
\vspace{1mm}
\noindent\begin{tabular}{ll}
\bf Title   & \textbf{SpeakEasy} \\
\bf Course  & CS/QTM/LING-329: Computational Linguistics \\
\bf Authors & Hyunbae Jeon, BS in Computer Science, \texttt{harry.jeon@emory.edu} \\ 
            & Rhea Ramachandran, BS in Computer Science, \texttt{rhea.ramachandran@emory.edu} \\
            & Victoria Ploerer, BA in Computer Science BS in Neuroscience, \texttt{vploere@emory.edu} \\
            & Yella Diekmann, BS in Computer Science, \texttt{ydiekma@emory.edu} \\
            & Max Bagga, BA in Computer Science, \texttt{max.bagga@emory.edu}
            
\end{tabular}
\vspace{1mm}
\hrule


\section*{Abstract}

Social interactions and conversation skills separate the successful from the rest and the confident from the shy. For college students in particular, the ability to converse can be an outlet for the stress and anxiety experienced on a daily basis along with a foundation for all-important career skills. In light of this, we designed SpeakEasy: a chatbot with some degree of intelligence that provides feedback to the user on their ability to engage in free-form conversations with the chatbot. SpeakEasy attempts to help college students improve their communication skills by engaging in a seven-minute spoken conversation with the user, analyzing the user's responses with metrics designed based on previous psychology and linguistics research, and providing feedback to the user on how they can improve their conversational ability. To simulate natural conversation, SpeakEasy converses with the user on a wide assortment of topics that two people meeting for the first time might discuss: travel, sports, and entertainment. Unlike most other chatbots with the goal of improving conversation skills, SpeakEasy actually records the user speaking, transcribes the audio into tokens, and uses macros—e.g., sequences that calculate the pace of speech, determine if the user has an over-reliance on certain words, and identifies awkward transitions—to evaluate the quality of the conversation. Based on the evaluation, SpeakEasy provides elaborate feedback on how the user can improve their conversations. In turn, SpeakEasy updates its algorithms based on a series of questions that the user responds to regarding SpeakEasy's performance. The source code for SpeakEasy can be found here: \href{https://github.com/HarryJeon24/SpeakEasy.git}{https://github.com/HarryJeon24/SpeakEasy.git}. See README for detailed instructions about setup. 


\section{Team Vision}
\label{sec:team-vision}

Communication is essential in all facets of life. The way people talk, express themselves, and share ideas is at the forefront of our daily lives. Humans are a social species, which is why communicating effectively is key to maintaining relationships, getting a job, and increasing self-confidence. For instance, 95.9\% of employers found written/oral communication essential for jobs, but estimated that a mere 41.6\% of employees actually possessed this skill~\cite{nace-2018-jobs} Furthermore, research suggests that improving communication and social skills can help alleviate development of social anxiety~\cite{https://doi.org/10.1016/j.jaac.2017.01.007}. This is where SpeakEasy comes into play.

\subsection{Overall Objective}
\label{ssec:overall-objective}
The goal of SpeakEasy is to help people improve their communication skills through researched-backed metrics and tips. A dialogue system is perfect to achieve this goal, as it provides the perfect avenue for the user to practice and receive feedback on their communication without the social pressure of interacting with another human being. SpeakEasy first engages in a friendly conversation with the user. On top of providing great practice for small talk in a safe environment, this dialogue is used to collect information about how the user communicates and which areas can be strengthened. Such information can only be collected through a dialogue system. Then, the chatbot presents constructive feedback that the user can easily implement into future conversations. After using SpeakEasy, users will be prepared to build stronger relationships, land and keep jobs more easily, and build their confidence. As such, the overarching goal of SpeakEasy is to improve the user's overall quality of life.


\subsection{Target Audience}
\label{ssec:target-audience}

The target audience for our system is college students, both males, and females. We decided to cater to this age group because communication skills are essential for personal growth and development. However, many individuals struggle to express themselves due to social anxiety or other speaking disorders. This has especially been heightened through the pandemic with 88\% of students suffering from stress, 44\% from anxiety, and 36 \% from severe depression \cite{Lee2021-zf}. While an inability to communicate can be attributed to these disorders, \citet{Lee2021-zf} claims that lack of communication can also be one of the determinants of the aforementioned disorders. In other words, the inability to communicate can cause anxiety and other disorders which in turn worsens the student's communicative skills: a vicious cycle. With this in mind, we believed we optimized SpeakEasy's impact by designing it for college students. 

An individual's experience in college often determines the course of the rest of their academic and personal lives. Thus, by catering our system to college students, we can help them become more confident in themselves and their abilities as they navigate different relationships, job opportunities, and personal growth. Individuals who can communicate effectively can express their feelings, needs, and ideas clearly. This, in turn, can improve their relationships and overall well-being. Studies evaluating students before, during, and after university experience has established college as the most profound time in a person's life in which they can learn cooperative skills \cite{10.3389/fpsyg.2018.01536}. If learned successfully, this cooperative methodology serves as a building block from which future corporate and office skills develop. Consequently, it is imperative that college students successfully develop the communication skills necessary during the short time they have in university and as a result, SpeakEasy likely has its most predominant impact when used by college students.

We specifically catered to university students by incorporating conversational topics that appeal to college students. Although we directed the conversation to engage with college students, everyone can benefit from interacting with our system. The ability to make conversion is a universal skill that people continue to improve on throughout the course of their lives.


\section{Challenges}
\label{sec:challenges}
The main challenge with developing SpeakEasy was creating a chatbot that understood questions the user may ask. With the nature of SpeakEasy, it is expected that the user will try to have an engaging conversation since they know their conversational abilities will be judged. The best way to increase engagement is to ask questions -- it is one of the quantitative metrics measured, after all. However, predicting every possible question the user may ask and generating a response is near impossible. To overcome this challenge, SpeakEasy has opinions and a personality. For example, this chatbot wants to live in Barcelona and loves the beach. Adding depth to SpeakEasy allows it to answer personal questions the user may ask, increasing engagement and interactivity. However, there may still be instances where the user's questions go unanswered. To guarantee that these instances were not negatively impacting the user experience too much, one aspect of the evaluation included how natural the conversation felt. Ideally, SpeakEasy is able to acknowledge what the user is saying and answer enough questions to resemble real human conversation.

Furthermore, understanding human emotion became a challenge as well. To ensure that our conversational AI operated fast enough to simulate as much of a proper conversation as possible, we attempted to minimize the amount of GPT API calls. Consequently, we could not always flag the emotion of the user's input using the GPT API. Unfortunately, this results in some instances where SpeakEasy responds positively to something clearly negative which would not happen in an actual conversation and actually detracts from the natural nature of the conversation. We attempted to circumnavigate this by designing a dialogue flow that could have more neutral ts response. Because the entire aim of our project is to help the user improve their ability to converse in a realistic environment, maintaining the realism of our conversation was the most pertinent challenge. With this in mind, we attempted to ensure that the flow of the conversation was as seamless as possible: i.e., no abrupt transitions to other topics or off-topic dialogue; however, since we cannot account for all possible user inputs, ensuring that this was the case was no easy task. To try to prevent against making abrupt transitions, we tried to preempt transitions to other topics by priming the user with questions that could generate user responses that relate to the topic SpeakEasy intends to switch to. 

A technical challenge arose when using Natex with the speech to text technology. These two technologies are incompatible, so regular expressions embedded within Macros were used to overcome this. Other technical challenges came from the differential operating systems of Windows and Mac. For example, the string inputs for the systems calls accessing the operating system and thus the speakers differ for each version. Consequently, we had to develop different versions for each kind of system. Additionally, package installation is far more difficult on Mac as certain packages require complicated procedures in order to properly install the right version. Background noise also played a huge technical challenge. If they system cannot properly pick up the audio from the background, then the string input for the GPT API calls can often passed in garbled text—e.g., if it is noisy in the background, sometimes characters and words from languages are picked up which force the GPT to throw errors which forces it to terminate early. As a result, we tried to ensure that some of the error statements force the dialogue flow to ask further questions as opposed to simply terminating the code. Finally, SpeakEasy relies on the GPT API to process what the user is saying. Though this increases its language understanding capabilities immensely, the GPT API experiences major slowdowns at times which negatively impacts user experience. Thankfully, the use of an API key mitigated this negative effect.


\section{Dialogue Overview}
\label{sec:dialogue-overview}

\begin{figure}[htp]
    \centering
    \includegraphics[width=14cm]{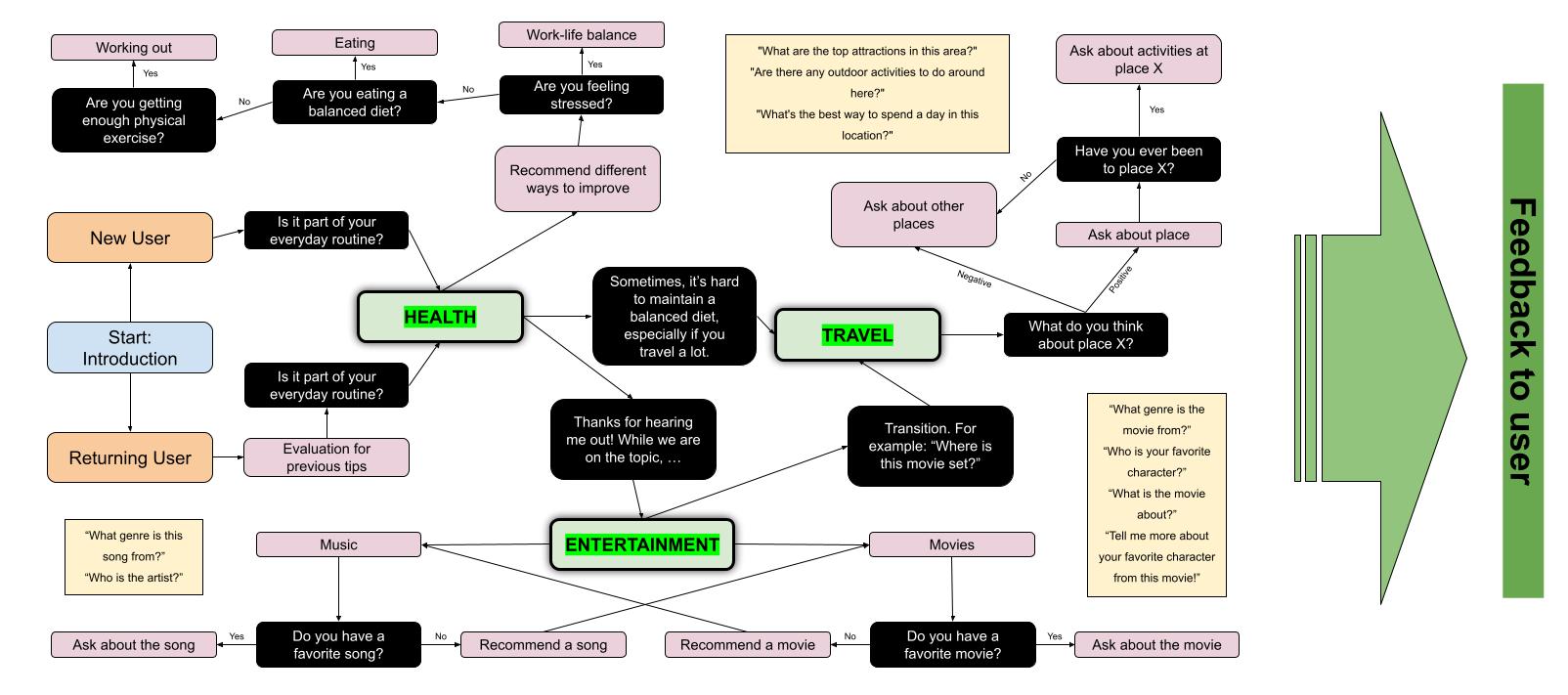}
    \caption{A diagram of the dialogue flow}
    \label{fig:dialogue_flow}
\end{figure}

The chatbot's dialogue flow creates a natural conversation that lasts for at least seven minutes (averages around ten minutes) and covers topics related to travel, health, and entertainment. The general idea is that two strangers meet for the first time and create small talk. Therefore, the dialogue will start with questions and phrases such as "Hello, what is your name?", "My name is…", "How are you doing?", "I am doing…". In the case that the person is a returning user, the system conducts a different greeting that incorporates a short evaluation of the user's prior experience with the bot. After the introduction, the chatbot guides the conversation into talking about health. We wanted to design the flow such that the user does not feel forced to talk about the topics; therefore, the chatbot uses the guiding questions “Is it part of your everyday routine?” and “Do you make it part of your daily schedule?” to transition into talking about health. During the conversation about health, the chatbot asks about and discusses the user’s eating habits, workout routine, and work-life balance. Following these topics, the chatbot stirs the conversation either towards talking about entertainment or about travel. The latter transition, for example, is done by pointing out the difficulties of maintaining a healthy diet while traveling. 

During the travel conversation, the chatbot asks the user questions such as “Where are you from?”, "Where would you rather live?" and "Are there any other places you would like to live in or visit?". Furthermore, the chatbot asks the user for recommendations on what to do in specific travel locations and, if they prefer, would ever consider solo travel. The chatbot is also able to share information about its persona. For example, the chatbot is able to tell the user that it is from Austria and that it wants to live in Barcelona. At the end of this conversation, the chatbot enters the feedback state of the conversation.

When talking about entertainment, the chatbot is able to converse about movies and music. It first asks the user if they have a favorite movie or song, and if so, ask questions about it. For example, “Who is your favorite artist for that type of music?”, “What genre is it”?, and “Who is your favorite character in the movie?”. The chatbot then moves on to recommending a song or movie to the user. If the user is satisfied, the chatbot either transitions over to talking about travel or it will end the conversation and enter the feedback state of the conversation. The travel state can transition back over to the movie state in the event that a concept from a movie is brought up—e.g., did you know that this movie was filmed here—in order to prevent the conversation from being fully linear. 

During the feedback state, the chatbot provides information about the user's conversation performance and previously determined metrics, such as how many questions the user asks, word choice, talking speed, and the level of attention that the user gives to their conversation partner. It also asks the user if it wants to hear the underlying metrics of the analysis: e.g., the specific statistics we incorporated from previous research in order to evaluate the conversation.

Lastly, the chatbot asks the user for feedback on its performance. The user is be able to do so by answering questions like “Did I say anything wrong during the conversation?” and “Did you ever have a hard time understanding me?" in order to assess the chatbots performance. These questions were chosen in line to the evidence presented in section \ref{sec:evaluation}.

\begin{figure}[htp]
    \centering
    \includegraphics[width=12cm]{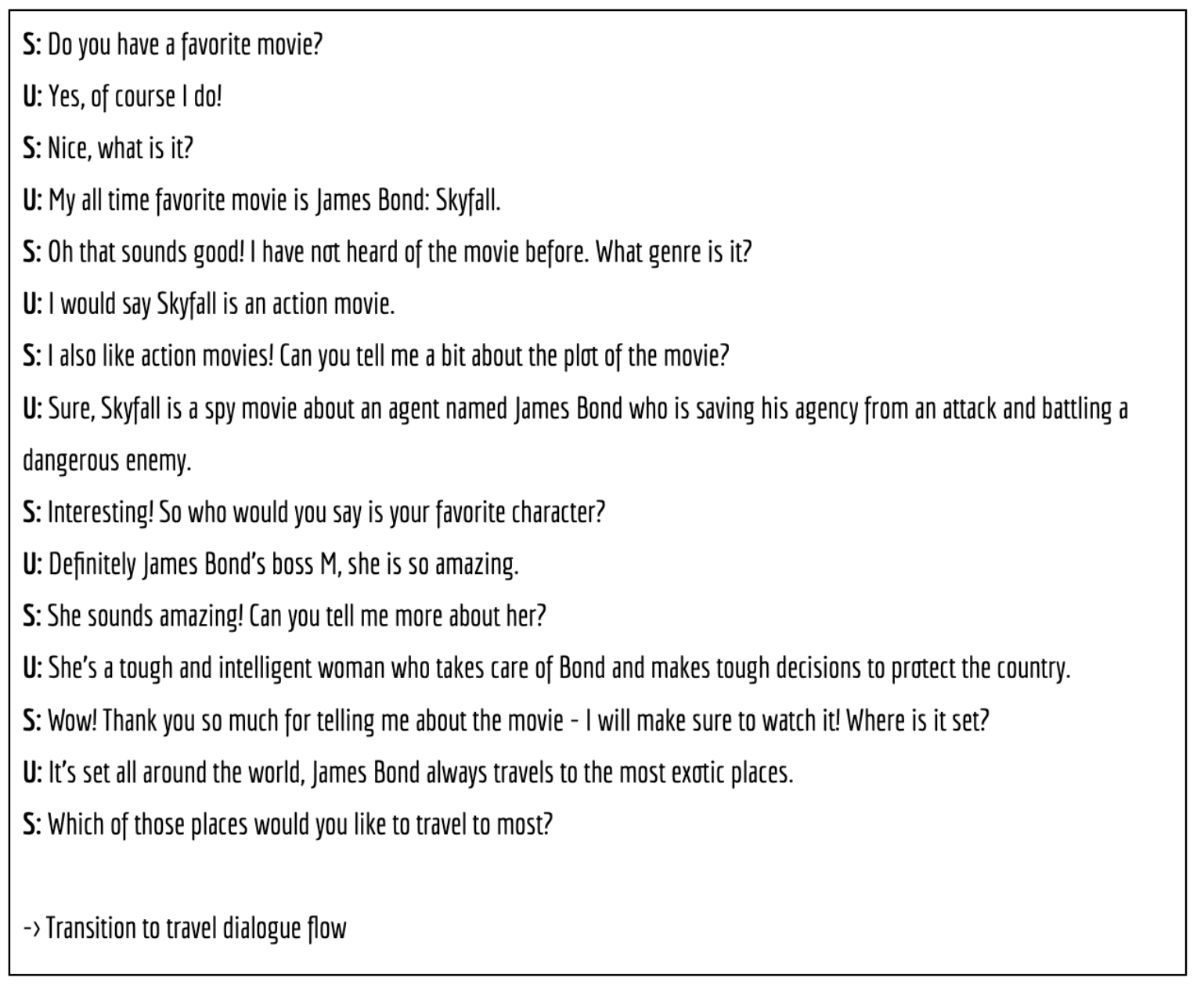}
    \caption{Sample dialogue conducted by the chatbot}
    \label{fig:dialogue_flow-sample}
\end{figure}

Figure \ref{fig:dialogue_flow-sample} provides a sample dialogue conducted by the chatbot during the entertainment dialogue flow. Because the user has a favorite movie, the chatbot proceeds to ask questions about the movie. After some time, the chatbot ends the conversation and proceed to giving feedback to the user about their performance.

\section{Methodology}
For the conversation framework, we are using EmoraSTDM \cite{finch-choi-2020-emora}, which allows developers to add custom macros and ontologies. One of the built-in features in EmoraSTDM is Natex, which functions similarly to regular expressions. However, we chose not to utilize Natex in our implementation. Instead, we relied on the GPT API \cite{openai2023gpt4} to understand user input. It is important to note that we did not use GPT to generate responses in this system. To improve conversational skills, we aimed to mimic a human conversation environment as closely as possible. Given that more people experience anxiety when talking rather than texting \cite{doi:10.1089/cpb.2006.9936}, we decided to implement speech-to-text and text-to-speech macros.

\subsection{Audio Function}

In this project, we aimed to develop a chatbot capable of taking audio input and producing audio output. To achieve this, we utilized OpenAI's Whisper Automatic Speech Recognition (ASR) API \cite{https://doi.org/10.48550/arxiv.2212.04356} for transcribing audio to text and the gTTS (Google Text-to-Speech) package in Python for converting text to audio. This section details the scientific approach and methods employed in creating the chatbot. To enable the chatbot to accept audio input, we used the PyAudio library in Python to record the user's voice. The following parameters were used for recording:

\begin{itemize}
    \item Format: 16-bit integer (paInt16)
    \item Channels: 1 (mono)
    \item Sample rate: 44,100 Hz
    \item Chunk size: 1,024 samples
\end{itemize}

The audio recording was saved as a WAV file, and a separate thread was created to handle the recording process. This allowed users to stop the recording by pressing the Enter key. The duration of the recording was calculated by measuring the time elapsed between the start and end of the recording process. After obtaining the audio input, we utilized OpenAI's Whisper API to transcribe the audio to text. The API was called with the Whisper ASR model "whisper-1" and the recorded audio file. The API returned a transcript of the audio, which was then assigned to a variable for further processing. For text-to-audio conversion, we employed the gTTS package in Python. The gTTS library converts text to audio using Google's Text-to-Speech API. To generate audio output, we passed the chatbot's response text to the gTTS function, specifying the following parameters:

\begin{itemize}
    \item Text: The chatbot's response
    \item Language: English (en)
\end{itemize}

The gTTS function saved the synthesized audio as an MP3 file. To play the audio output, we used the "os.system()" function in Python. For Windows, we executed the "start" command, followed by the MP3 file's name, and used the "time.sleep()" function to pause the execution of the script for the duration of the audio playback, which was determined using the MP3 library to extract the length of the audio file. For macOS, we executed the "afplay" command, followed by the MP3 file's name, without the need for a sleep function.
The chatbot's workflow was as follows:

\begin{enumerate}
    \item Record the user's audio input using the MacroRecordAudio class.
    \item Transcribe the audio input to text using OpenAI's Whisper API.
    \item Process the text input and generate a response using the chatbot's logic.
    \item Convert the text response to audio using the MacrogTTS class.
    \item Play the audio output for the user.
\end{enumerate}

This approach allowed us to create a chatbot capable of interacting with users through audio input and output, offering a more natural and accessible user experience.

\subsection{Transitions}
While creating the introduction state, we used GPT to get the user's name. From there we store all user names as a key in the USERS dictionary which we use to check whether the user has previously used the system. If they are a returning user, we store their rating for their previous interaction with the bot as well as the rest of their feedback later in the dialogue flow. We also use GPT for retrieving information like how the user is doing and what they have been up to. In addition to this, we use a variety of GET and SET macros to store GPT responses in various variables which we can use to retrieve the information later through our .pkl file.

For the Health transition, we again utilize GPT as well as several GET and SET macros to store information about the user's health (i.e. the user's lifestyle, whether or not they exercise, their eating habits, etc). We also utilized IF macros to determine whether or not a user indicated yes or no to a particular question. 

The Entertainment section works in a similar way, using GTP and GET/SET macros to save user information. It also utilizes the Spotify API for song suggestions and the The Movie Database (TMDB) API for movie recommendations.

Once again, GTP and macros was utilized to build the Travel portion of the conversation. SpeakEasy prompts the user to talk about where they are from and where they have traveled or want to travel. Beyond this, SpeakEasy can discuss favorite activities at these locations and the user's specific travel habits. To be able to acknowledge what the user is saying, many SET and GET macros were employed to store and retrieve responses. GTP was only used to process what the user is saying, not to generate responses. Beyond this, SpeakEasy's personality shines through in the travel portion. We made this chatbot have a home country, and even specific travel wishes. 

\subsection{Feedback}

A set of macros is designed to help SpeakEasy provide advice to users by evaluating their conversational skills. We focus on five aspects for evaluation: awkward transition, verbal tic, acknowledgment, number of questions, and number of words per minute. This evaluation matrix is based on three papers \cite{doi:10.1080/08856257.2018.1458472}, \cite{doi:10.2466/pr0.102.1.111-118}, \cite{doi:10.1080/08824099409359938}.

The work by \citet{doi:10.1080/08856257.2018.1458472} highlights the importance of intonation, smooth transition, and asking questions for good conversation. Based on these findings, we created the MacroAwkward, MacroNumQuestions, and MacroAVGToken classes. The MacroAwkward class checks for awkward transitions by comparing the user's transitional words to a predefined list of awkward transitional words, which were generated by ChatGPT for suggestion and can be expanded \cite{doi:10.1080/08856257.2018.1458472}. If the user uses less than 10 awkward transitions, they are considered to be doing well. The MacroNumQuestions class counts the number of question marks in user responses and advises the user to ask more questions if the ratio of questions to total utterances is less than 0.39, as this was the high question rate that showed positive effects on liking towards the conversation partner \cite{huang2017doesn}. The MacroAVGToken class measures the user's intonation by calculating the words per minute (WPM) and providing appropriate feedback based on a range of 120 to 150 WPM, which is the average American talking speed in conversation according to the cited source \cite{virtualspeechAverageSpeaking}.

Research by \citet{doi:10.2466/pr0.102.1.111-118} demonstrates the negative impact of overusing specific words. To address this aspect, we created the MacroTic class, which calculates the frequency of each token the user used, excluding common words like articles. The class provides feedback on the overuse of certain words based on the sorted frequency list of tokens.

Experiments conducted by \citet{doi:10.1080/08824099409359938} reveal the importance of acknowledging and showing empathy towards the speaking partner. To measure these elements, we implemented the MacroAcknow class. One way to show acknowledgment and empathy is by mentioning what the speaking partner said previously. The MacroAcknow class calculates the language style matching (LSM) using a function words list commonly used in conversation, comparing the user's responses to SpeakEasy's responses. If the LSM is greater than or equal to 0.8, the user is considered to have successfully shown attention to their partner under symmetric, cooperative, social conditions, where the two people had to accomplish a task. This is in contrast to asymmetric, competitive, or negotiation conditions, where the LSM may not hold the same impact \cite{richardson2019cooperation}. Otherwise, they are advised to show more attention by acknowledging what their partner said previously.

\section{Evaluation}
\label{sec:evaluation}

Because for Quiz 6, we all designed separate evaluation plans, we chose one of the evaluation plan's statistics to present here and detail our reasoning behind our evaluation. Prior to defining the five categories for the evaluation, we gathered information about the diversity of the participants that evaluated our chatbot SpeakEasy. Each member was assigned to gather ten individuals to evaluate our system.  Of the ten total participants in the context of this evaluation plan, six of them were female and four of them were male. The age of the participants ranged from 20-22 while the average age was 21.2 years old. All ten participants came from middle to upper-middle class backgrounds. four of the participants were from Caucasian descent, five were from Asian descent, and one was from African American descent. The context of all ten conversations were casual, the purposes of the conversations were to improve conversational skills, and the power dynamics of the conversation were that all participants were my peers.

The metric chosen was the Likert scale from 1-5. In class, we went over the Likert scale from 1-3; however, we decided to expand it to 1-5 because according to one of the original papers using the Likert scale, the optimal metrics were (1) Strong Disagree, (2) Disagree, (3) Undecided, (4) Agree, and (5) Strongly Agree. \citet{likert-susan} argues that the extra feeling of intensity (e.g., between disagree and strongly disagree) is important and because it allows for better statistical analysis. Furthermore, according to their analysis, the difference in intensity between each category is roughly even allowing for a well-distributed interval for the variable. 

In terms of satisfaction in a conversation with a chatbot, according to \citet{Mohelska-eval}, the main attributes that must be fulfilled is if the chatbot met expectation and left an impression. Consequently, the two questions we asked were “did the chatbot met expectations” and “did the chatbot leave an impression”. All ten participants gave a five to both these questions; however, from a qualitative perspective, it was less the dialogue flow and conversation, and more the voice technology that stunned the user. As a result, in the future, evaluations of conversation satisfaction should ask questions specific to dialogue flow. \citet{Mohelska-eval} also recommends asking about the conversation duration and number of conversation turns which we did. All but one participant strongly agreed that the conversation duration was appropriate while the other user said they agreed that the conversation duration was appropriate. All ten people evaluating SpeakEasy said that the number of conversation turns was appropriate. 

According to \citet{Vanjani-eval}, the biggest limitation in evaluation techniques was evaluating how natural the chatbots are. They argue that chatbots can be conversationally unnatural in two ways: the chatbot can give garbled responses or give too accurate of an answer. As a result, we decided to ask, “did the chatbot give garbled responses” and “were the answers the chatbot gave were unnaturally accurate”. The answers for whether the chatbot gave garbled responses ranged from 2-4 with the average answer being 2.8. The biggest reasons for why the participants responded undecided or disagree were that the chatbot failed to respond accurately to questions or completely ignored. This is something that the chatbot needs to improve on going forward. Specifically, we can use the GPT API to first determine whether is asking a question, then use the GPT API to determine if the user is being on topic, and finally randomly select a response from an answer bank that we construct. 

When generating automatic conversation evaluators, \citet{Yi-eval} determined that the four most determiners of chatbot coherence were whether the system response is comprehensible, on topic, interesting, and if the user wanted to continue the conversation. As a result, we ask four questions: (1) “Are SpeakEasy’s responses comprehensible”, (2) “Are SpeakEasy’s responses on topic”, (3) “Are SpeakEasy’s response interesting”, and (4) “Did you want to continue the conversation”. The average response for comprehensibility was a 4.8. The average response for on topic was a 3.8 with the answers ranging from a 2-5. The main qualitative issues were the aforementioned cases in which SpeakEasy fails to respond to questions and in certain cases SpeakEasy switches topics with fully closing out the previous topic. The average response for whether the conversation was interesting was a 4.2. Participants that did not give a strongly agree mainly stated that while sometimes SpeakEasy gave a response specific to the participant’s response, there are some instances in which it could give a more user-centric response. All five participants that responded 5 to whether they wanted to continue the conversation gave qualitative feedback mostly stating that they loved being able to have a conversation in which they could actually speak to the chatbot. 

When we designed SpeakEasy, we wanted it to be as empathetic as possible because we felt that a person cannot be good at conversation without introducing some degree in empathy into the conversation. The main question asked by a study evaluating chatbot empathy conducted by \citet{agarwal-eval} was whether the chatbot disregarded any of the feelings of the user. As a result, we asked the following question “Did SpeakEasy disregard any emotion you interjected into the conversation” and “Did SpeakEasy reciprocate any emotions that you conveyed”. For the former question, results varied from 1-5 with the average response being 3.1. A particularly concerning qualitative response was that in response to the user saying that they were crying, SpeakEasy responded to “I enjoy crying”. We must fix this in the future as this error can drive away potential users. To the later question, the average response was a 3.8 with no specific qualitative concerns. 

\citet{wade-eval} argue that the interpret-ability of conversational AI depends on a variety of factors most importantly if the chatbot can explain its reasoning, its ability to contain facts that make sense to the user, and to want unwanted bias. Therefore, we asked the following questions: (1) “was the chatbot’s reasoning behind its feedback well-explained”, (2) “were the facts presented in the feedback clear”, and (3) “was there any bias in the analysis”. All users but one gave a 5 for the first question, with the one dissenting participant giving it a 4 stating that the LSM score was unclear. An explanation of this score can be explained better going forward. For questions 2 and 3, all users gave a 5 saying that the feedback was accurate and easily understood. 

Based on the aforementioned analyses, we need to improve on few specific areas. Most importantly, we need to redesign SpeakEasy to be able to recognize and answer questions better. A more minor upgrade would be better designing our chatbot to recognize specifics of the user response and changing states (the number of which would increase) to accordingly respond to the user. Specifically, we need to recognize negative responses better and avoid responding in a positive tone as that can appear condescending and scare off users. Finally, we should explain the LSM score better and in future evaluations we should ask questions more explicitly about the dialogue flow.


\section{Novelty}
\label{sec:novelty}

During free conversation with SpeakEasy, the chatbot saves the user's and the bot's conversation logs. These logs are then used to evaluate and provide feedback on the user's conversational skills, focusing on five aspects: awkward transition, verbal tic, acknowledgment, number of questions, and number of words per minute. This evaluation matrix is based on three papers \cite{doi:10.1080/08856257.2018.1458472}, \cite{doi:10.2466/pr0.102.1.111-118}, \cite{doi:10.1080/08824099409359938}.

The work by \citet{doi:10.1080/08856257.2018.1458472} highlights the importance of intonation, smooth transition, and asking questions for good conversation. Based on these findings, we created the MacroAwkward, MacroNumQuestions, and MacroAVGToken classes. The MacroAwkward class checks for awkward transitions by comparing the user's transitional words to a predefined list of awkward transitional words, which were generated by ChatGPT for suggestion and can be expanded \cite{doi:10.1080/08856257.2018.1458472}. If the user uses less than 10 awkward transitions, they are considered to be doing well. The MacroNumQuestions class counts the number of question marks in user responses and advises the user to ask more questions if the ratio of questions to total utterances is less than 0.39, as this was the high question rate that showed positive effects on liking towards the conversation partner \cite{huang2017doesn}. The MacroAVGToken class measures the user's intonation by calculating the words per minute (WPM) and providing appropriate feedback based on a range of 120 to 150 WPM, which is the average American talking speed in conversation according to the cited source \cite{virtualspeechAverageSpeaking}.

Research by \citet{doi:10.2466/pr0.102.1.111-118} demonstrates the negative impact of overusing specific words. To address this aspect, we created the MacroTic class, which calculates the frequency of each token the user used, excluding common words like articles. The class provides feedback on the overuse of certain words based on the sorted frequency list of tokens.

Experiments conducted by \citet{doi:10.1080/08824099409359938} reveal the importance of acknowledging and showing empathy towards the speaking partner. To measure these elements, we implemented the MacroAcknow class. One way to show acknowledgment and empathy is by mentioning what the speaking partner said previously. The MacroAcknow class calculates the language style matching (LSM) using a function words list commonly used in conversation, comparing the user's responses to SpeakEasy's responses. If the LSM is greater than or equal to 0.8, the user is considered to have successfully shown attention to their partner under symmetric, cooperative, social conditions, where the two people had to accomplish a task. This is in contrast to asymmetric, competitive, or negotiation conditions, where the LSM may not hold the same impact \cite{richardson2019cooperation}. Otherwise, they are advised to show more attention by acknowledging what their partner said previously.

The idea of improving communication through technology is not new. However, there is a clear market gap
for the use dialogue systems to provide personalized feedback on conversation. There are two big competitors
that share similarities with SpeakEasy, yet neither provides the same service.
Rocky.AI is a personal development coaching app that provides help on improving communication.
However, its main purpose is to help the user use self-awareness and reflection to reach their personal
goals. Communication is just a small portion of this. On the other hand, SpeakEasy provides much more
in-depth feedback and is able to narrow down exactly what the user needs to work on without relying on
self-assessment questions. Rather, our metrics are based on research and use a quantitative approach to
assessing communication skills. Furthermore, SpeakEasy provides the user with the opportunity to practice
small-talk, something Rocky.AI lacks. Rocky.AI is a self-help coach app that focuses on self-reflection,
while SpeakEasy is a dialogue system that provides personalized feedback in hopes of improving the user’s
communication. Though both systems aim to improve communication, the approach Rocky.AI takes is
entirely different from what SpeakEasy aims to do.

Yoodli is our most similar competitor. It is an AI speech coach that records the user talking while they
are speaking in a professional setting, and then uses this information to provide personalized feedback on how
to the user can communicate more effectively. SpeakEasy and Yoodli share the same formulaic approach to
generate feedback. That is, both systems assess real dialogue the user has to give appropriate feedback. The
biggest difference is that Yoodli focuses on improving professional speech, while SpeakEasy aims to improve
communication of all types. While superficially this difference may be minute, it actually drives the novelty
of SpeakEasy. Presentations, which Yoodli specializes in, relies more on capturing the audiences attention by
changing the tone, volume, and pace of the speaking voice \cite{richardson2019cooperation}. In other words, the content of the speaking is
less important in public speaking: captivating the audience trumps all. SpeakEasy will generally improve
all forms of communication, but focuses mainly on conversations. Engaging conversations do not revolve
around the quality of speech but instead build of the creativity of the talk \cite{doi:10.1080/08856257.2018.1458472}. As a result, SpeakEasy focuses heavily on developing extensive ontology’s to capture a wide array of conversations and then perform a series
of analyses in order to analyze the quality of the subjects discussed. This skill itself is more important than
the public speaking skills developed by Yoodli with conversations being essential to mental health \cite{doi:10.1089/cpb.2006.9936}. In
short, SpeakEasy assesses communication in a casual conversation setting, while Yoodli provides feedback on
interviews and public speaking.


\section{Contributions}
\label{sec:contributions}

The development of SpeakEasy can be essentially divided into three main processes: development of the input audio and transcribing process, construction of the analysis Macros, and developing each of the main conversational transition topics. However, in order to abide by the central scrum tenants of having working software at the end of each sprint and not working on the software one step of time to avoid integration errors, we worked on several component simultaneously with different members working on different elements. Hyunbae Jeon and Max Bagga were primarily responsible for developing the speech to text process and developing the analysis macros. Hyunbae developed the windows version of the software while Max developed the Mac version of the software. Hyunbae contributed primarily by developing the initial framework for the audio input and output and developing the analysis Macros. Max worked primarily by refining the audio component, integrating the dialogue flows, and working on the supplemental Macros needed to make the dialogue flow work. Victoria Ploerer, Rhea Ramachandran, and Yella Diekmann developed the conversational transitions. Specifically, Rhea created the introduction and health transitions, while Victoria completed the Travel transition, and Yella created the Entertainment transition. Upon completing each transition, we all worked together to connect them in order to carry out a smooth transition from one topic to another. Despite the given roles, everyone was very involved in all aspects, and each of us contributed 20 percent to guarantee the chatbot functioned the best it can.

\bibliographystyle{plainnat}
\bibliography{proposal}

\end{document}